\documentclass[12pt]{iopart}

\usepackage{graphicx}

\begin{document}

\title{A bouncing ball model with two nonlinearities: a prototype for
Fermi acceleration}
\author{Edson D.\ Leonel and Mario Roberto Silva}

\address{Departamento de Estat\'{i}stica, Matem\'atica Aplicada e
Computa\c c\~ao -- IGCE -- Universidade Estadual Paulista -- UNESP --
Av. 24A, 1515 -- Bela Vista -- 13.506-900 -- Rio Claro -- SP -- Brazil}

\begin{abstract}
Some dynamical properties of a bouncing ball model under the presence
of an external force modeled by two nonlinear terms are studied. The
description of the model is made by use of a two dimensional nonlinear
measure preserving map on the variables velocity of the particle and
time. We show that raising the straight of a control parameter which
controls one of the nonlinearities, the positive Lyapunov exponent
decreases in the average and suffers abrupt changes. We also show that
for a specific range of control parameters, the model exhibits the
phenomenon of Fermi acceleration. The explanation of both behaviours
is given in terms of the shape of the external force and due to a
discontinuity of the moving wall's velocity.
\end{abstract}



\section{Introduction}
The bouncing ball model consists of a classical particle of mass $m$
which is confined to bounce between two infinitely heavy and
rigid walls \cite{ref1}. One of the walls is assumed to be fixed while
the other one moves in time according to a periodic function. This
model, also known as the Fermi-Ulam model (FUM), is a simple dynamical
system that can be modeled using the formalism of discrete mappings.
Moreover, many tools developed to characterise such a model show to
have great applicability in more complex mappings. Considering the
FUM, many results are known in the literature. Particularly for the
particle suffering elastic collisions with either walls, it is known
that the phase space of the model is of mixed kind \cite{ref2} in the
sense that depending on the combination of both the control parameters
and initial conditions, invariant spanning curves limiting the size of
chaotic seas and Kolmogorov-Arnold-Moser (KAM) islands can all be
observed. The presence of the invariant spanning curves yields in a
limit for the energy gain of a bouncing particle, thus the Fermi
acceleration (unlimited energy gain of the particle) is not observed.
A similar version of the model, called as bouncer \cite{pustil},
consists of a classical particle, in the presence of a constant
gravitational field, suffering elastic collisions with a periodically
moving platform. The returning mechanism of the bouncer model, a
mechanism that injects the particle for a next collision with the
moving wall, is rather distinct of the FUM. In the bouncer, it is due
only to the gravitational field while in the FUM it is given by a
collision with a fixed wall. These differences yield in a profound
consequence for the dynamics of a bouncing particle. Depending on the
control parameter, the unlimited energy gain is observed in the bouncer
model, a phenomenon that is not present in FUM with periodic and smooth
oscillations. The differences were clarified by Lichtenberg, Lieberman
and Cohen \cite{ref7}. Hybrid versions of the FUM and bouncer were
recently studied \cite{ref13,ref9} as well as a stochastic version of
the FUM \cite{karlis}.

There are also many important results concerning the inclusion of
damping forces on both the models (see for example Ref. \cite{ref11}
for a short review). One of them is the presence of a drag force
\cite{bb1}, so that the particle is moving inside a gas with the
dissipation acting on the particle along its trajectory. The dynamics
of the problem is, generally, given by a nonlinear mapping that is
obtained via the solution of Newton's law. A different kind of
dissipation can be introduced via inelastic hits of the particle with
the walls. Thus, there is a restitution coefficient that makes the
particle experiences a fractional loss of energy upon collisions.
Despite both
kinds of damping often occur in nature, they have profound
and different consequences in the dynamics of the models. As an
example, in Refs. \cite{ADD7,ADD8} and considering inelastic
collisions, Tsang and Lieberman considered the simplified FUM (both the
walls are fixed but the particle changes energy and momentum upon
collisions with one of the wall as if the wall were moving) with
inelastic impacts. They have evidenced contraction on the phase space
and in particular, observed the presence of a strange attractor.
Recently, a rather similar version of the dissipative model
\cite{crisis},
confirmed the property of area contraction and in
addition, a boundary crisis was characterised. Additionally, a family
of boundary crisis was observed when collisions with the two walls are
inelastic \cite{pla}. The bouncer
model was also considered under
inelastic collisions. For example, in \cite{ADD3} Holmes discusses the
appearances of horseshoes in the inelastic bouncer and gave an
illustration of a homoclinic orbit in such a model. After that, Everson
\cite{new1} presents and discusses
with many numerical simulations the
appearance of period doubling
cascade in the damping bouncer model.
Period doubling cascade was also observed in \cite{ADD4} for the
completely inelastic collisions. The presence of frictional force
however was considered by Luna-Acosta \cite{ADD6} and Naylor, Sanch\'ez
and Swift \cite{ADD5} in the bouncer model. They too observed period
doubling cascades and in special Luna-Acosta \cite{ADD6} has achieved
analytically dimensional reduction for the limit of high dissipation.

In this letter, we study a non dissipative version of a bouncing
ball model seeking to understand and describe some of its dynamical
properties considering however that the motion of the moving wall is
given via a crank-connecting rod scheme. For such a scheme, it is known
that there are two nonlinearities present in the model and each of them
play important rules in the dynamics. Depending on certain ranges
of control parameters, there can be profound consequences on the
dynamics of the system. Particularly, when one of the two control
parameters is raised, the positive Lyapunov exponent experiences a
drastic reduction. It is also important to say that the particle is in
the total absence of any external field. Other important result for
this model is that it yields, for specific control parameter values,
the phenomenon of Fermi acceleration (unlimited energy growth). The
phenomenon is characterised, for the first time in the present model,
in terms of a discontinuity of the derivative of the wall's position
with respect to the time, thus leading the particle to acquire
unlimited energy gain.

This paper is organised as follows. In section \ref{sec2} we present
the model and the expressions of the mapping that fully describes the
dynamics of the system. Section \ref{sec2} is also devoted to a
discussion of the numerical results and the behaviour of the positive
Lyapunov exponent. In section \ref{sec3} we propose a simplified
version of the model and study the behaviour of the average velocity
as function of a control parameter. We show that Fermi acceleration
emerges naturally from the deterministic dynamics of the model for
specific control parameter values. Final remarks and conclusions are
drawn in section \ref{sec4}.

\section{The model and the mapping}
\label{sec2}

\begin{figure}[htb]
\centerline{\includegraphics[width=0.8\linewidth]{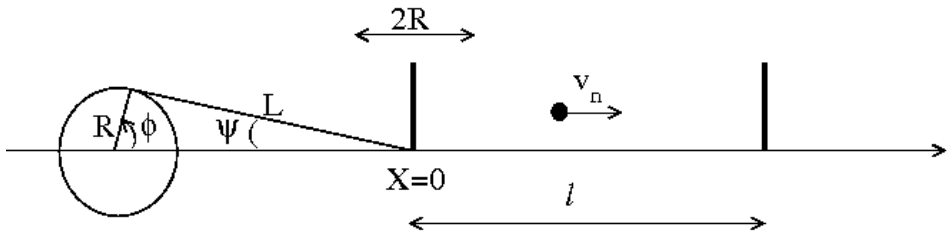}}
\caption{Illustration of the model under consideration.}
\label{fig1}
\end{figure}

The model is described using a two dimensional mapping for the
variables $(v_n,t_n)$, where $v_n$ and $t_n$ are the corresponding
velocity of the particle and time immediately after the n$^{th}$
collision with the moving wall. We assume that one wall is fixed at
$x=l$ and that the motion of the moving wall is given by
$s(t)=R\cos(wt)+\sqrt{L^2-R^2\sin^2(wt)}$ (we stress the term $wt$
in the equation of $s(t)$ is represented by the variable $\phi$ in Fig.
\ref{fig1}), where $R$ denotes the radii of the crank, $L$ is the
length of the connecting rod and $w$ is the corresponding frequency of
oscillation. Figure \ref{fig1} illustrates the model under
consideration. We stress the equation for $s(t)$ is easily obtained
from the condition of $R\sin(\phi)=L\sin(\psi)$, as can be seen in Fig.
\ref{fig1}.
Before we write the equations of the mapping, let us first discuss the
initial conditions. We assume that, at a time $t=t_n$, the particle
is at the position $x_p(t_n)=s(t_n)$ with velocity $v=v_n>0$. Thus such
an initial condition can be considered as if the dynamics were already
running in the system along the time. We emphasise that two different
kinds of collisions can be observed namely: (i) multiple hits
with the moving wall and (ii) a single hit with the moving wall. In
case (i), the particle suffers a collision with the moving wall but
then, before it leaves the collision zone, which is defined as
$x\in[-R,R]$, the particle experiences a second and then successive
impact with the moving wall. Such kind of collisions becomes rare in
the limit of high energy but they are quite often to be observed in the
regime of low energy. It is also easy to see that there are too many
control parameters in the model, 4 in total, namely $R$, $L$, $l$ and
$w$ and that the dynamics of the system does not depend on all of them.
It is convenient to define dimensionless and more appropriated
variables. We define $\epsilon=R/l$, $r=R/L$, $V_n=v_n/(wl)$ and
measure the time in terms of the number of oscillations of the moving
wall $\phi_n=w t_n$. For the dimensionless variables, we consider that
the range for $\epsilon$ is $\epsilon\in[0,1]$ and for $r$ is
$r\in[0,1]$. The limit of $r\rightarrow 0$ corresponds to
$L\rightarrow\infty$ and $r\rightarrow 1$ is obtained for
$L\rightarrow R^+$. With this new set of variables, the mapping that
describes the dynamics of the system is written as
\begin{equation}
T:\left\{\begin{array}{ll}
\phi_{n+1}=[\phi_n+\Delta T_n]~~{\rm mod (2\pi)}\\
V_{n+1}=V_n^*-2\epsilon\sin(\phi_{n+1})\left[1+{{r\cos(\phi_{n+1})}
\over { \sqrt { 1-r^2\sin^ 2(\phi_{n+1})}}}\right]\\
\end{array}
\right.,
\label{eq1}
\end{equation}
where the expressions of $V_n^*$ and $\Delta T_n$ depend on the kind
of collision. For case (i), which corresponds to the multiple hits
with the moving wall, the corresponding expressions are $V_n^*=-V_n$
and $\Delta T_n=\phi_c$ with $\phi_c$ obtained by the solution of
$G(\phi_c)=0$ with $G(\phi_c)$ given by
\begin{eqnarray}
G(\phi_c)&=&\epsilon\cos(\phi_n+\phi_c)-\epsilon\cos(\phi_n)-V_n\phi_c+
{{\epsilon}\over{r}}\sqrt{1-r^2\sin^2(\phi_n+\phi_c)}-\\ \nonumber
&-&{{\epsilon}\over{r}}\sqrt{1-r^2\sin^2(\phi_n)}~.
\label{eq2}
\end{eqnarray}
A solution of the function $G(\phi_c)$ for $\phi_c\in(0,2\pi]$
corresponds to a collision of the particle with the moving wall and it
is obtained numerically.

Let us now consider the case where the particle leaves the collision
zone, i.e. case (ii). The corresponding expressions are $V_n^*=V_n$,
$\Delta T_n=\phi_T+\phi_c$ where $\phi_T$ corresponds to the elapsed
time the particle spends travelling from the last hit with the
moving wall, up to suffering an elastic reflection with the static wall
and be reflected backwards, therefore until the entrance of the moving
wall. Thus, $\phi_T$ is given by
\begin{equation}
\phi_T={{2+\left({{\epsilon}\over{r}}-{{\epsilon}\over{r}}\sqrt{
1-r^2\sin^2(\phi_n)}
\right)-\epsilon\cos(\phi_n)-\epsilon}\over{V_n}}~.
\label{eq3}
\end{equation}
The term $\phi_c$ is numerically obtained from $F(\phi_c)=0$ for
$\phi_c\in[0,2\pi)$ where the function $F(\phi_c)$ is given by
\begin{equation}
F(\phi_c)=\epsilon\cos(\phi_n+\phi_T+\phi_c)+{{\epsilon}\over{r}}\sqrt{
1-r^2\sin^2(\phi_n+\phi_T+\phi_c)}-{{\epsilon}\over{r}}
-\epsilon+V_n\phi_c~.
\label{eq4}
\end{equation}
After some straightforward algebra, it is easy to show that the mapping
(\ref{eq1}) preserves the following phase space measure
\begin{equation}
d\mu=\left[V+\epsilon\sin(\phi)\left(1+{{r\cos(\phi)}\over{\sqrt{
1-r^2\sin^2(\phi)}}}\right)\right]dVd\phi~.
\label{eq5}
\end{equation}
We stress that in the limit of $r\rightarrow 0$, the results for the
one-dimensional Fermi accelerator model are all recovered
\cite{ref13,ref12}.

\begin{figure}[t]
\centerline{\includegraphics[width=0.7\linewidth]{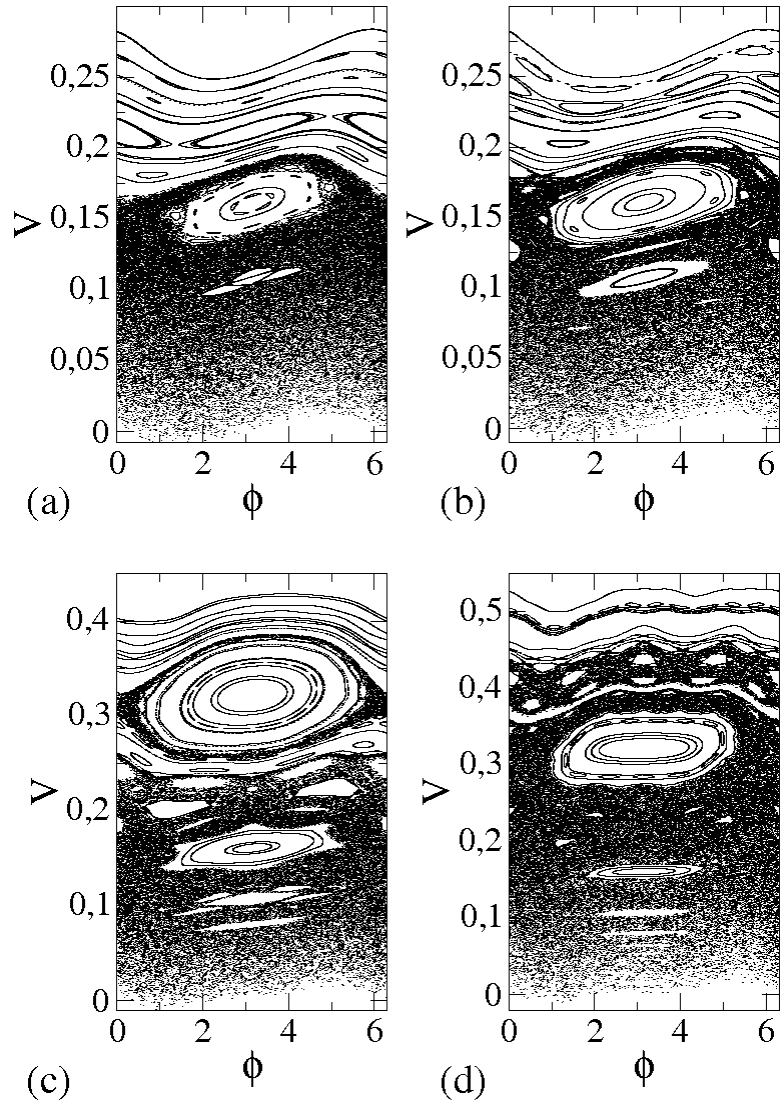}}
\caption{Phase space generated from mapping (\ref{eq1}) and control
parameter $\epsilon=0.01$ and: (a) $r=0.1$, (b) $r=0.3$, (c) $r=0.6$
and (d) $r=0.9$.}
\label{fig2}
\end{figure}

Figure \ref{fig2} shows
the corresponding phase space obtained via iteration of mapping
(\ref{eq1}) for the control parameter $\epsilon=0.01$ and: (a) $r=0.1$,
(b) $r=0.3$, (c) $r=0.6$ and (d) $r=0.9$.  We can clearly see that the
shape of the phase space changes as the control parameter $r$ varies.
On the other hand, the mixed form is preserved in the sense that a
large chaotic sea, which surrounds KAM islands, is limited by a set of
invariant spanning curves. One can also note that the position of the
lowest invariant spanning curve raises as the control parameter $r$
increases. In our simulations we will consider values for $r$ that may
approaches the unity, moreover in the range of $r\in[0,1]$. However for
a real experimental system, in which damping forces can not be
neglected, such values for $r$ have no much interest. This is mainly
because the damping force can acquires larger values as compared to the
component of the force with respect to the motion (for instance, it
happens for large values of the angle $\psi$) and therefore, lead the
system to reach the rest.

The two natural questions that we are interested in concern on the
properties of the chaotic sea (see Fig. \ref{fig2}), like the positive
Lyapunov exponent and the average velocity of the particle. Our main
goal is to describe their behaviour as function of the control
parameter $r$. We think this study is of interest because the control
parameter $r$ directly controls the straight of a nonlinearity of the
model. For small values of $r$, results of the FUM should be obtained.
Moreover, we expect that the results obtained for $r\rightarrow 1$
contribute towards a better understanding of this model for such a
range of $r$ and, in particular as we will see in Sec. \ref{sec3}, in a
description of the phenomenon of Fermi acceleration. The behaviour of
the Lyapunov exponent is described in this section while the average
velocity is discussed in section \ref{sec3}.

\begin{figure}[t]
\centerline{\includegraphics[width=0.5\linewidth]{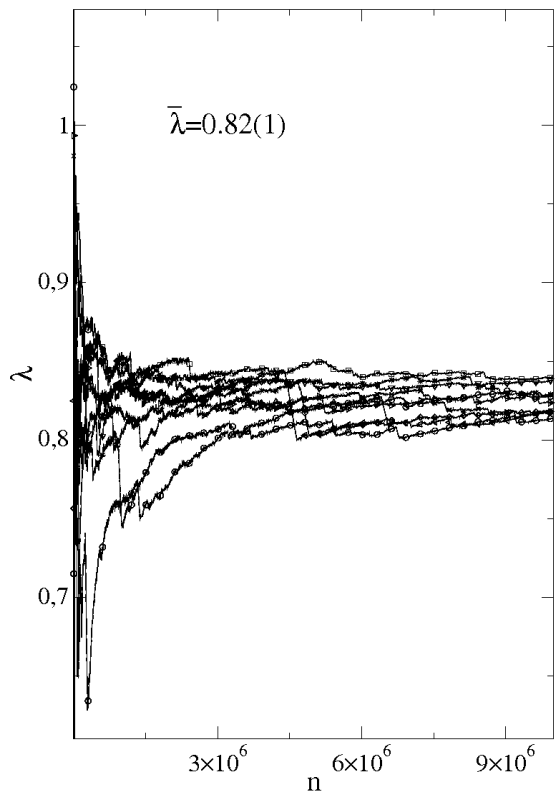}}
\caption{Positive Lyapunov exponent for 10 different initial
conditions randomly chosen along the chaotic sea in the low energy
regime. The control parameters used were $\epsilon=0.01$ and $r=0.1$.}
\label{fig3}
\end{figure}

We concentrate to investigate the behaviour of the positive Lyapunov
exponent for the chaotic sea. It is well known that the Lyapunov
exponent is commonly used as a tool to characterise sensitivity to
initial conditions. Figure \ref{fig3}
shows the asymptotic convergence of the positive Lyapunov exponent for
the control parameters $\epsilon=0.01$ and $r=0.1$. The ensemble
average of $10$ different initial conditions randomly chosen along the
chaotic sea gives $\bar \lambda=0.82\pm 0.01$, where the error $0.01$
denotes the standard deviation of the ten samples. Each initial
condition was iterated up to $10^7$ collisions with the moving wall.
The method used to obtain the Lyapunov exponents is described in the
Appendix.

\begin{figure}[t]
\centerline{\includegraphics[width=0.5\linewidth]{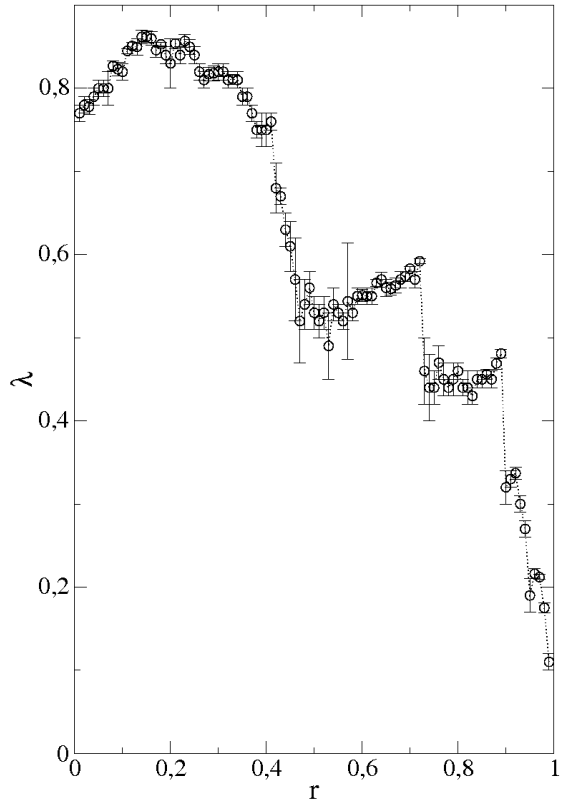}}
\caption{Positive Lyapunov exponent as function of $r$ for a fixed
$\epsilon=0.01$.}
\label{fig4}
\end{figure}

Let us discuss the behaviour of the positive Lyapunov exponent as
function of $r$. Figure \ref{fig4} shows
the behaviour of $\lambda\times r$. We can see that, in the limit of
$r\rightarrow 0$, the positive Lyapunov exponent recovers the value of
the one-dimensional Fermi accelerator model \cite{ref13,ref12}. The
positive Lyapunov exponent then grows slightly, having a maximum value
around $r=0.15$ and then decreases almost monotonically until around
$r=0.5$. Then it starts grow again until $r\approx 0.7$ when it
suddenly decreases. Other abrupt change is observed for $r\approx
0.89$. Thus, the two main questions that arise from Fig. \ref{fig4} 
are: (i) why does the positive Lyapunov exponent decreases in the
average, instead of growth, as $r$ raises? (ii) what is the
explanation of the abrupt changes in $\lambda$?

\begin{figure}[t]
\centerline{\includegraphics[width=0.6\linewidth]{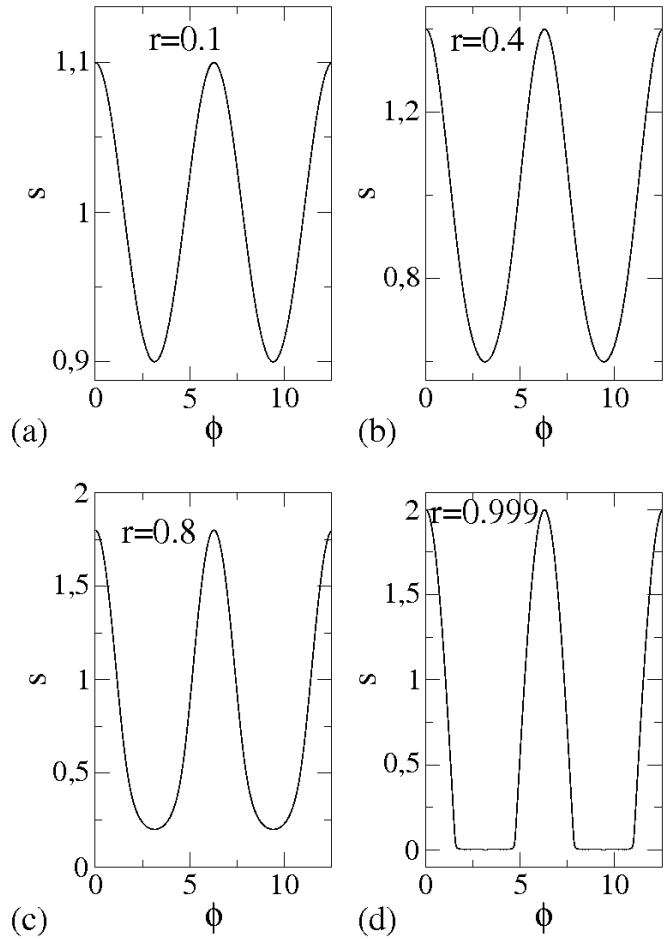}}
\caption{Plot of $S(\phi)$ for two periods in $\phi$ considering the
control parameters $\epsilon=0.01$ and: (a) $r=0.1$, (b) $r=0.4$, (c)
$r=0.8$ and (d) $r=0.999$.}
\label{fig5}
\end{figure}

The answer for question (i) comes from the shape of the function
that describes the motion of the moving wall. Figure \ref{fig5} shows
four different plots of $S(\phi)=r\cos(\phi)+\sqrt{1-r^2\sin^2(\phi)}$
(for sake of clarity, we show two periods in $\phi$ for
$S(\phi)$), which describes the motion of the moving wall for four
different values of $r$ namely: (a) $r=0.1$, (b) $r=0.4$, (c) $r=0.8$
and (d) $r=0.999$. For $r=0.1$, the function looks like a cosine
function but as $r$ increases, the shape changes substantially. It thus
become to have two regimes of variation where one of them is
characterised by a constant plateau in the limit of $r\rightarrow 1$,
as can be seen in the range of $\phi\in(\pi/2,3\pi/2)$ and a regime of
fast variation, which is given by the complementary values of $\phi$.
In the limit case of $r=1$, the function $S(\phi)$ has indeed
discontinuities in its derivative for two values of $\phi$, namely
$\phi=\pi/2$ and $\phi=3\pi/2$. As we will see in the next section,
the discontinuities for $S(\phi)$ yield a profound consequence in the
dynamics of the system thus leading the particle to exhibit unlimited
energy growth. The large plateaus imply that the particle, when suffers
collisions with the moving wall does not change substantially its
velocity value since such plateaus lead in an ``almost'' null velocity
for the moving wall. Thus the time that the particle spends until a
next hit with the moving wall is almost the same as it spent in the
previous collisions. It then implies that the particle, in a chaotic
orbit, can experience many more collisions with the moving wall without
substantially changing its energy as it would experiences if the
plateaus were absent. Then, the form of $S(\phi)$ for large values of
$r$ yields in reducing the chaoticity of the system in the average.

\begin{figure}[t]
\centerline{\includegraphics[width=0.5\linewidth]{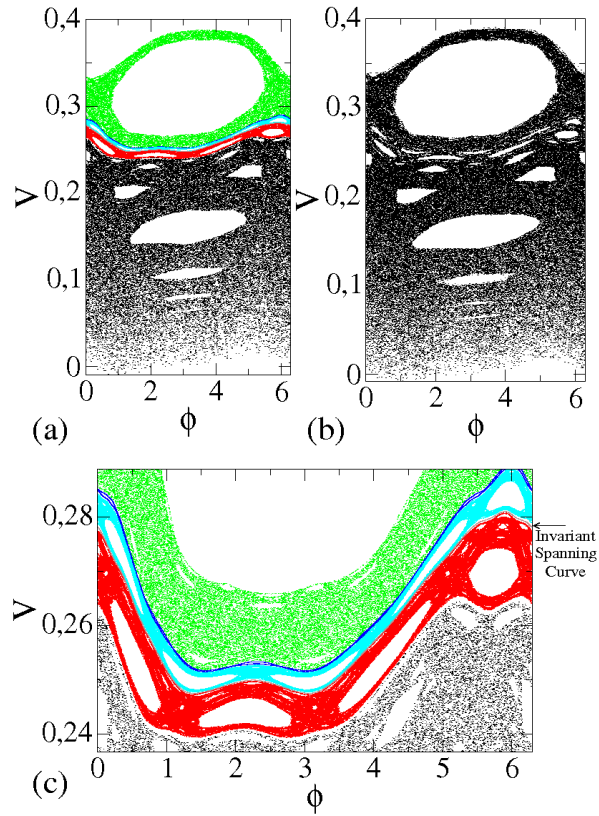}}
\caption{(a) Illustration of chaotic behaviour generated by $5$
different initial conditions for the control parameters
$\epsilon=0.01$ and $r=0.69$. (b) Iteration of a single initial
condition, evidencing the chaotic behaviour, for $\epsilon=0.01$
and $r=0.72$. (c) Zoom in of part (a) where it is easy to see an
invariant spanning curve and chaotic layers.}
\label{fig6}
\end{figure}

We will now discuss a possible answer for question (ii), i.e., the
explanation of the sudden changes in $\lambda$. They are basically
related to the destruction of the lowest energy invariant spanning
curve together with a destruction of small chaotic layers. Considering
the case of low energy and for $r=0.69$, the system has a large chaotic
sea (the black region of Fig. \ref{fig6}(c)) which shares boundary
with a thin chaotic layer (the red region of Fig. \ref{fig6}(c)), as
can be seen in Fig. \ref{fig6}.
Above this chaotic layer, there is an invariant spanning curve
(brown curve of Fig. \ref{fig6}(c)). Above yet of such curve we can
see two thin chaotic layers (light blue and dark blue) and a relatively
large chaotic sea (green region) surrounding a KAM island (see for
instance Fig. \ref{fig2}(c)). Each of these chaotic regions were
characterised in terms of Lyapunov exponents. Their corresponding
values were: for the black region we obtained $\lambda_{\rm
b}=0.574(6)$; for the red region $\lambda_{\rm
r}=0.072(1)$; for the light blue $\lambda_{\rm lb}=0.0110(4)$; for the
dark blue $\lambda_{\rm db}=0.0152(2)$ and for the green region
$\lambda_{\rm g}=0.218(1)$. For $r=0.72$, which is quite
close to $r=0.69$, those regions shown in Fig. \ref{fig6}(a)
were all merged into a single and large chaotic sea characterised by a
positive Lyapunov exponent $\lambda=0.46(5)$. The merged regions are
shown in Fig. \ref{fig6}(b). After the destruction of the thin
structures shown in Fig. \ref{fig6}(a), the chaotic sea in the low
energy region can spreads over a larger accessible region of the phase
space. So we can consider that, after the transition, the positive
Lyapunov exponent could be obtained by an average of the previous
values for the corresponding chaotic regions therefore taking into
account the fraction of area occupied individually by each region. To
check whether this supposition is correct, we have obtained the
fraction of each chaotic region previous to the control parameter
variation i.e. for $r=0.69$. We have defined a grid of $84$ initial
conditions for the $\phi$-axis, limited to the interval
$\phi\in[0,2\pi)$, and $127$ for the $V$-axis considering the interval
$V\in[0,0.38813]$. The value $0.38813$ corresponds to the higher value
of the velocity obtained for the chaotic sea shown in the green region
of Fig. \ref{fig6}(c). Thus in the total, we considered $10668$
different initial conditions. Each of them were evolved in time for
$n=10^7$ collisions with the moving wall and their Lyapunov exponents
evaluated. The obtained value was compared to the Lyapunov exponent of
the chaotic regions so that we were able to compare the corresponding
fraction of initial conditions which belongs to one region or to
another one. Applying this procedure for all the chaotic regions shown
in Fig. \ref{fig6}(c), we found that the black chaotic sea fills a
fraction of $P_{\rm b}=0.8261$ of the entire chaotic region. The red
region corresponds to $P_{\rm r}=0.0279$, when the light blue has a
fraction of $P_{\rm lb}=0.0103$, the dark blue is $P_{\rm db}=0.0038$
and finally the green region corresponds to a fraction of $P_{\rm
g}=0.1319$. After the transition, we could assume that the positive
Lyapunov exponent is obtained by
\begin{equation}
{\bar\lambda}=\lambda_{\rm b}P_{\rm b}+\lambda_{\rm r}P_{\rm
r}+\lambda_{\rm lb}P_{\rm lb}+\lambda_{\rm db}P_{\rm db}+\lambda_{\rm
g}P_{\rm g}~.
\label{eqnew}
\end{equation}
Evaluating Eq. (\ref{eqnew}) we found ${\bar\lambda}=0.50$ which is a
rather acceptable value as compared to the value obtained via numerical
simulation of the chaotic sea after the destruction $\lambda=0.46(5)$.

The abrupt change in $\lambda$ around the value of $r\approx 0.89$ is
explained by using the same arguments. We stress that similar results
were observed in a rather distinct model \cite{poco}.

\section{A simplified version of the model and Fermi acceleration}
\label{sec3}

Other important conclusion that arises from the shape of the function
$S(\phi)$ is related to the energy gain of the bouncing particle. As
$r$ approaches the unity, the variation of the moving wall position
becomes more fast for specific ranges of $\phi$. It implies that the
particle can acquires, for specific ranges of $\phi$, large values of
velocity upon collisions with the moving wall for those regions of fast
variation of $S(\phi)$. Such a result can be seen in Fig. \ref{fig2} by
the position of the lowest energy invariant spanning curve which assume
higher values as the control parameter $r$ raises. To illustrate such
an argument more clearly, it is important to look at the behaviour of
the average velocity for sufficiently long time. To do this, we will
make use of a simplification in the mapping (\ref{eq1}) with the main
goal of speeding up our numerical simulations and avoid solving the
equations $G(\phi_c)=0$ and $F(\phi_c)=0$. This simplification, which
is commonly used in the literature (see Refs.
\cite{ref1,ref13,karlis,ref16,cc1,cc2}), consists in assume that both
walls are fixed. However, when the particle hits one of them, it
exchanges energy and momentum as if the wall were moving. This
procedure retains the nonlinearity of the problem and yields a huge
advantage of avoid solving transcendental equations. The mapping is
then given by
\begin{equation}
T:\left\{\begin{array}{ll}
\phi_{n+1}=[\phi_n+{{2}\over{V_n}}]~~{\rm mod (2\pi)}\\
V_{n+1}=\left|V_n-2\epsilon\sin(\phi_{n+1})\left[1+{{r\cos(\phi_{n+1})}
\over { \sqrt { 1-r^2\sin^ 2(\phi_{n+1})}}}\right]\right|\\
\end{array}
\right..
\label{eq6}
\end{equation}
Although this simplification brings the advantage of allowing very fast
simulations as compared to those of the complete version, it also gives
rise to a problem that we need to avoid. In the complete model,
depending on the combination of both velocity and phase of the
moving wall, it is possible for the particle, after suffering a
collision with the time varying wall, to suffer a second successive
collision before exiting the collision area, as well as possibly
having a negative velocity following the first such a collision. In the
simplified model, non-positive velocities are forbidden because they
are equivalent to the particle travelling beyond the wall. In order to
avoid such problems, if after the collision the particle has a
negative velocity, we inject it back with the same modulus of velocity.
Such a procedure is effected perfectly by the use of a modulus
function. Note that the velocity of the particle is reversed by the
modulus function only if, after the collision, the particle remains
travelling in the negative direction. The modulus function has no
effect
on the motion of the particle if it moves in the positive direction
after the collision. We stress that this approximation is valid only
for small values of $\epsilon$.

We now discuss the procedure used to obtain the average velocity of
the particle. Firstly we obtain the average velocity proceeding with
an average over the orbit, i.e.
\begin{equation}
V_i={{1}\over{n}}\sum_{j=1}^nV_{i,j}~,
\label{eq7}
\end{equation}
where $j$ denotes the number of collisions. The second average is
made in an ensemble of $M=10^3$ different initial conditions so that
the average value is defined as
\begin{equation}
{\bar{V}}={{1}\over{M}}\sum_{i=1}^MV_i~,
\label{eq8}
\end{equation}
with $i$ corresponding to a sample in an ensemble of $M$ different
initial conditions.

\begin{figure}[t]
\centerline{\includegraphics[width=0.45\linewidth]{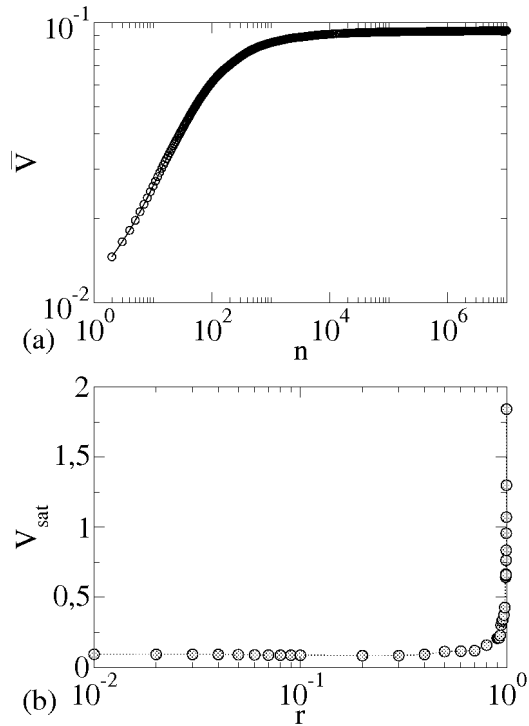}}
\caption{(a) Behaviour of ${\bar V}\times n$ for the
control parameters $r=0.01$ and $\epsilon=0.01$. (b) Plot of
$V_{\rm sat}\times r$ for a fixed $\epsilon=0.01$.}
\label{fig7}
\end{figure}

\begin{figure}[t]
\centerline{\includegraphics[width=0.45\linewidth]{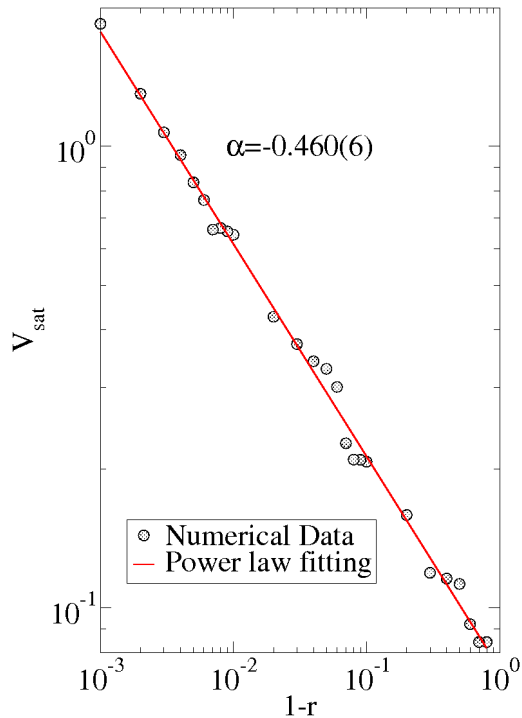}}
\caption{Plot of $V_{\rm sat}\times \mu$ for a fixed $\epsilon=0.01$. A
power law fitting gives that $\alpha=-0.460(6)\cong -0.5$.}
\label{fig8}
\end{figure}

Figure \ref{fig7}(a) shows
the behaviour of ${\bar{V}}\times n$ for the control parameters
$\epsilon=0.01$ and $r=0.01$ considering $n=10^7$ iterations. We can
see that the velocity of the particle grows for short iterations and
then suddenly it bends towards a regime of saturation for sufficiently
long time. Moreover, we are interested in the behaviour of the
asymptotic values of ${\bar{V}}$. Figure \ref{fig7}(b) shows the
behaviour of ${\bar V}$ for long iterations, which we will referr to it
as $V_{\rm sat}$. The behaviour of $V_{\rm sat}$ can be described for
two different ranges of $r$. The first range is for $r<0.9$ where we
can see that $V_{\rm sat}$ is almost constant for the region of
$r\in[0,0.9)$. The second range of $r$ is considered for $r\ge0.9$. It
is important to say that, when $r=1$ the expression of the ``moving
wall'' velocity presents discontinuities for $\phi=\pi/2$ and
$\phi=3\pi/2$. Thus, it is convenient to define a new parameter
$\mu=1-r$ and then study the behaviour of $V_{\rm sat}$ as a function
of $\mu$. This new control parameter has a practical application 
because it brings the criticality of the model to $\mu=0$. Figure
\ref{fig8} shows
the behaviour of $V_{\rm sat}\times \mu$ for a fixed control parameter
$\epsilon=0.01$. We can describe such a behaviour according to
\begin{equation}
V_{\rm sat}\propto \mu^{\alpha}~.
\label{eq9}
\end{equation}
After fitting a power law in Fig. \ref{fig8}, we obtain
$\alpha=-0.460(6)\cong -0.5$. This kind of behaviour for $V_{\rm sat}$
confirms that, in the limit of $r\rightarrow 1$, the present model
shows the phenomenon of Fermi acceleration. This result has a
clear explanation in terms of the KAM theorem. As it was discussed by
Lichtenberg, Lieberman and Cohen \cite{ref7}, if the expression of the
periodic wall's velocity has a sufficient number of continuous
derivatives, then it is possible to obtain invariant spanning curves
separating different portions of the phase space. Particularly they are
useful to prevent unlimited energy growth for a bouncing particle.
Moreover, it was estimated by Moser \cite{ref18} that three continuous
derivatives for the moving wall velocity is a sufficient condition for
the existence of KAM surfaces (by instance, invariant spanning curves).
However, in the present model and considering the limit of
$r\rightarrow 1$, the expression of the velocity shows discontinuities
for $\phi=\pi/2$ and $\phi=3\pi/2$, so that the invariant spanning
curves are not observed for $r=1$ and consequently, Fermi acceleration
is present.

\section{Conclusions}
\label{sec4}
In summary, we have studied a non dissipative version of a classical
bouncing ball model under the presence of two nonlinearities. Our
results show that, as one of the two control parameters varies, the
positive Lyapunov exponent diminish in the average and experiences
sudden changes. We have explained such a behaviour by the shape of the
moving wall and due to the destruction of invariant spanning curves
and thin chaotic layers. We have also shown that in the limit of $r
\rightarrow 1$, the present model exhibits unlimited energy growth.
This phenomenon was explained by using a discontinuity of the moving
wall's velocity.

\section*{Acknowledgements}
EDL thanks Prof. Tadashi Yokoyama for fruitful discussions. Support
from CNPq, FAPESP and FUNDUNESP, Brazilian agencies is gratefully
acknowledged.

\section*{Appendix}

In  this section, we briefly discuss the procedure used to obtain the
Lyapunov exponents. In effect, the procedure consists in evolving the
system over a long time from two slightly different initial conditions.
If the two trajectories diverge exponentially in time the orbit is
called chaotic, and the Lyapunov exponent obtained is positive. If
the Lyapunov exponent is negative, the orbit may be either periodic or
quasi-periodic. Let us now describe the procedure used to obtain the
Lyapunov exponents numerically. They are defined as \cite{ref14}
$$
\lambda_j=\lim_{n\rightarrow \infty} \ln{|\Lambda_j|},~~~~~~~j=1,2,
$$
where $\Lambda_j$ are the eigenvalues of
$M=\prod_{k=1}^nJ_k(V_k,\phi_k)$ and $J_k$ is the Jacobian matrix
evaluated over the orbit $(V_k,\phi_k)$. The Jacobian matrix is
defined as
$$
J=\left(\begin{array}{ll}
{{\partial V_{n+1}}\over{\partial V_n}}  &  {{\partial
V_{n+1}}\over{\partial \phi_n}}  \\
{{\partial \phi_{n+1}}\over{\partial V_n}}  &    {{\partial
\phi_{n+1}}\over{\partial \phi_n}}\\
\end{array}
\right).
$$
In order to evaluate the eigenvalues of $M$, we use the fact that $J$
can be written as a product of $J=\Theta T$, where $\Theta$ is an
orthogonal matrix and $T$ is a right upper triangular one. We now
define the elements of these matrices as
$$
\Theta=\left(\begin{array}{ll}
\cos(\theta)  &   -\sin(\theta) \\
\sin(\theta)  &~~    \cos(\theta)\\
\end{array}
\right),~~ T=\left(\begin{array}{ll}
T_{11}  &   T_{12} \\
0  &  T_{22}\\
\end{array}
\right).
$$
Since $M$ is defined as $M=J_nJ_{n-1}\ldots J_2J_1$, we can introduce
the identity operator, rewrite $M$ as $M=J_nJ_{n-1}\ldots
J_2\Theta_1\Theta_1^{-1}J_1$, and define $\Theta_1^{-1}J_1=T_1$. The
product $J_2\Theta_1$ defines a new matrix $J_2^*$. In a following
step, we may write $M$ as $M=J_nJ_{n-1}\ldots
J_3\Theta_2\Theta_2^{-1}J_2^*T_1$. The same procedure yields
$T_2=\Theta_2^{-1}J_2^*$. The problem is thus reduced to the
evaluation of the diagonal elements of $T_i:T_{11}^i,T_{22}^i$. Using
the $\Theta$ and $T$ matrices, we find the eigenvalues of $M$, given
by
$$
T_{11}={{j_{11}^2+j_{21}^2}\over{\sqrt{j_{11}^2+j_{21}^2}}},~~
T_{22}={{j_{11}j_{22}-j_{12}j_{21}}\over{\sqrt{j_{11}^2+j_{21}^2}}}.
$$
We can then evaluate the Lyapunov exponent using the relation
$$
\lambda_j=\lim_{n\rightarrow
\infty}\sum_{k=1}^n{1\over{n}}\ln{|T^k_{jj}|},~~~~~~~j=1,2.
$$
It is interesting to observe that $\lambda_1=-\lambda_2$, because the
map is measure-preserving.

\end{document}